\documentclass[preprint,aps,prd,12pt,preprintnumbers,eqsecnum,nofootinbib,superscriptaddress]{revtex4-2}
\usepackage{xcolor}
\usepackage{braket}
\usepackage{amssymb,amsmath,amsfonts,comment,latexsym,graphicx,epstopdf,float}
\usepackage{color}
\usepackage{fancyvrb}
\usepackage{chngcntr}
\usepackage{etoolbox}
\usepackage{ulem}
\usepackage{array}
\usepackage{bbold}
\usepackage{subfigure}
\usepackage{slashed}
\usepackage{multirow}

\usepackage[colorlinks=true,citecolor=blue,linkcolor=red,filecolor=purple,urlcolor=magenta]{hyperref}

\usepackage{float}

\newcommand{\be}{\begin{equation}}
\newcommand{\ee}{\end{equation}}
\newcommand{\ba}{\begin{eqnarray}}
\newcommand{\ea}{\end{eqnarray}}

\newcommand{\grts}{\raise.3ex\hbox{$>$\kern-.75em\lower1ex\hbox{$\sim$}}}
\newcommand{\lets}{\raise.3ex\hbox{$<$\kern-.75em\lower1ex\hbox{$\sim$}}}

\usepackage{color}
\usepackage{amssymb,amsmath,amsfonts}
\usepackage{epstopdf} 
\usepackage{braket}

\usepackage{autobreak}

\begin{document}
%
%  
% Title of paper
\title{\vspace*{0.5in} 
Active Learning and Genetic Search for Discrete Abelian Flavor Model Building 
\vskip 0.1in}
\author{Selin A\c{s}mano\u{g}lu}\email[]{sasmanoglu@wm.edu}
%\affiliation{High Energy Theory Group, Department of Physics, William \& Mary, Williamsburg, VA 23187-8795, USA} 
\author{Christopher D. Carone}\email[]{cdcaro@wm.edu}
\affiliation{High Energy Theory Group, Department of Physics, William \& Mary, Williamsburg, VA 23187-8795, USA}
\date{November 18, 2026}
%\date{\today}
%
%
\begin{abstract}
The sequential breaking of discrete flavor symmetries provides a theoretical framework that can explain quark mass and mixing angle hierarchies. Each model is identified by its field content, charge assignments and symmetry breaking pattern, leading to an enormous space of possible models; this motivates the investigation of computational approaches for identifying promising candidates.  In this work, we focus on quark flavor models with the flavor symmetry group $G_F=(\mathbb{Z}_2)^p$, where $p$ is a positive integer; operators involving flavon fields which spontaneously break the flavor symmetry determine the size of the Yukawa matrix entries.  We consider two search strategies: an active learning search employing a random forest regressor and a genetic search. The generation of candidate models is automated and models are evaluated using a loss function that takes into account the logarithmic discrepancy between the model prediction and target observable values as well as penalties for model complexity. We find that both our active learning and genetic algorithms significantly outperform random model generation, which we use as a control; our active learning approach identifies lower-loss models faster (i.e., with fewer exact model evaluations) than our genetic search, though each ultimately finds successful models.  Our results support the effectiveness of active learning and evolutionary optimization as approaches to automate the search for promising models based on discrete flavor symmetries.
\end{abstract}
\pacs{}

\maketitle
\newpage

\section{Introduction} \label{sec:intro}

The spontaneous breaking of the Standard Model gauge symmetry gives rise to the quark and lepton masses. However, the resulting mass spectrum depends on the Yukawa couplings of the Higgs boson, whose values are not determined within the theory. Naively, one might expect the fermion masses to be comparable to the electroweak symmetry-breaking scale, but this expectation fails dramatically for all fermions except the top quark. The remaining quarks and leptons are substantially lighter. For example, the ratios of all the charged fermion masses to the top-quark mass range from unity down to ${\cal O}(10^{-6})$, exhibiting a striking hierarchical pattern that is not anticipated by naive dimensional analysis.

This mystery, known as the Standard Model flavor problem, can be addressed by models with flavor symmetries that act horizontally across the three Standard Model generations. For a flavor symmetry group $G_F$ that undergoes a sequential breaking to successively smaller subgroups,
\begin{equation}
G_F \stackrel{\langle \phi_1 \rangle}{\longrightarrow} H_1
\stackrel{\langle \phi_2 \rangle}{\longrightarrow} H_2
\stackrel{\langle \phi_3 \rangle}{\longrightarrow}
\cdots \rightarrow \mbox{nothing} \, ,
\label{eq:chain}
\end{equation}
small expansion parameters arise from the ratios of the vacuum expectation values (vevs) of the symmetry-breaking scalar fields, called flavons, $\phi_i$, to the ultraviolet cutoff of the effective flavor theory, $\Lambda_F$. The ratios $\langle \phi_i \rangle/\Lambda_F$ are hierarchical because the breaking of $G_F$ occurs at a sequence of decreasing energy scales. The resulting hierarchies can be further amplified in the Yukawa coupling matrices, whose entries arise from higher-dimensional operators involving products of flavon fields. Only the top-quark Yukawa coupling is required to remain invariant under $G_F$, allowing it to remain nonvanishing in the limit of unbroken flavor symmetry.

The construction of such models requires choosing the flavor symmetry group $G_F$, embedding the fermion fields into its irreducible representations, specifying a sequential symmetry-breaking pattern (i.e., the sequence of subgroups $H_i$), identifying the flavon fields $\phi_i$ and their quantum numbers that realize this breaking pattern, and selecting the energy scale associated with each stage of the symmetry breaking. This defines an enormous space of possible models. A significant part of the art of flavor model building lies in identifying those choices that yield models consistent with the observed fermion mass spectrum while remaining economical, aesthetically appealing, and, ideally, predictive. Consequently, the subject has generated a vast literature of proposed models (see, e.g., Refs.~\cite{Feruglio:2015jfa,Altmannshofer:2024hmr} for reviews), with no guarantee that the most compelling possibilities have yet been identified.

Given the vast landscape of flavor theories, it is natural to ask whether their exploration can be automated. The purpose of this paper is to make preliminary progress toward this goal, focusing on one particular class of models.

Computational approaches to flavor-model construction have appeared previously in several forms. Earlier bottom-up scans used algorithmic methods to identify discrete flavor symmetries consistent with desired lepton-mixing structures~\cite{Talbert:2014bda}. More recently, reinforcement-learning methods have been applied to Froggatt--Nielsen-type quark mass models~\cite{Harvey:2021oue}, to combined quark-and-lepton flavor models with a $U(1)$ flavor symmetry~\cite{Nishimura:2020nre}, and to autonomous construction of neutrino flavor theories~\cite{Baretz:2025zsv}. Machine-learning-inspired optimization techniques have also been used to identify neutrino mass-matrix textures with desired properties~\cite{Aranda:2026}, while neural-network-based methods have been developed to efficiently sample continuous parameter spaces in theories beyond the Standard Model~\cite{Hammad:2023}; see also Ref.~\cite{Giarnetti:2025mit} for a recent review.  The present work is complementary to these studies, focusing specifically on the systematic exploration of discrete flavor symmetry assignments and their associated Yukawa textures. We investigate two strategies for identifying phenomenologically promising models within this discrete theory space. The first employs active learning based on random-forest regression~\cite{Breiman:2001hzm} to predict the quality of candidate models and prioritize those warranting computationally expensive coefficient scans. The second uses a genetic algorithm~\cite{Holland:1975,Mitchell:1996} to evolve populations of candidate flavor models.

Specifically, we investigate whether machine learning (ML) can efficiently identify promising candidates within a class of discrete Abelian flavor models based on $(\mathbb{Z}_2)^p$ flavor symmetries. We first consider an active learning approach in which an ML algorithm is trained on an initial set of models to recognize promising candidates, and then applies what it has learned to identify additional good models within much larger collections of randomly generated models while continually refining its training. We also investigate a genetic approach in which pairs of promising models are combined to produce a new generation of ``offspring'' models that are better able to describe the fermion mass hierarchies. The terms ``good'' and ``improved'' are defined quantitatively through a loss function that is minimized when a model accurately reproduces a selected set of target observables.

Our goal is not to present a high-precision phenomenological analysis. For example, we do not use the latest experimental data and error bars or incorporate renormalization-group running, leaving such refinements to future work.  We also do not consider non-Abelian generalizations which can lead to more predictive and compelling models by relating multiple Yukawa matrix entries to a single operator coefficient. (For a review of non-Abelian discrete flavor models, see Ref.~\cite{Ishimori:2010au}; for a typical model of this type, see Refs~\cite{Aranda:1999kc,Aranda:2000tm}.)  Instead, we focus here on establishing the overall feasibility of our approach in a simplified setting, where the space of possible models can be readily encoded in a Python program.

Our paper is organized as follows. In Sec.~\ref{sec:framework}, we define the Abelian flavor models considered in this work and describe their computational representation. In Sec.~\ref{sec:ML}, we introduce the algorithmic methods used to explore this theory space, including an active-learning search based on random-forest regression and a genetic search. In Sec.~\ref{sec:results}, we show that both methods identify models with substantially lower loss than an unguided random scan using comparable numbers of exact model evaluations (defined in Sec.~\ref{sec:ML}), indicating improved agreement with the target flavor observables. We also consider the naturalness of the operator coefficients that result from the continuous optimization of the output models and we present examples of successful models that are identified by each of our search methods. Finally, in Sec.~\ref{sec:conc}, 
we summarize our conclusions and discuss directions for future work.

\section{Model Theory Space} \label{sec:framework}

In this section, we define the flavor model space over which the best model search is performed.  As noted earlier, discrete Abelian flavor symmetries have been of significant interest in the literature~\cite{Feruglio:2015jfa,Altmannshofer:2024hmr}.  We focus on a simple choice
\begin{equation}
G_F=(\mathbb{Z}_2)^p \,\,\, ,
\label{eq:GF}
\end{equation}
which represents $p$ repeated factors of the smallest discrete Abelian group, $\mathbb{Z}_2$, where $p$ is a positive integer.   This choice provides our search algorithms with a population of models to explore that range from those that are unlikely to be successful (for example, ones based on a single  $\mathbb{Z}_2$ factor and  one flavon field) to those that have substantial freedom to accommodate acceptable Yukawa hierarchies (for example, models with large values of $p$ and many flavon fields). Our goal is to use this model space to explore how effectively our search algorithms can do a task that is usually associated with human model builders:  identifying models that are successful in describing target observables and that are also as economical as possible in their field content and symmetry breaking pattern.   We work with a discrete global symmetry for this initial exploration; models with discrete gauge symmetries (i.e., discrete symmetries that also satisfy a set of anomaly cancellation conditions~\cite{Ibanez:1991hv,Banks:1991xj}) are also interesting, but will not be considered in the present work.    An added advantage of the choice in  Eq.~(\ref{eq:GF}) is that the charge assignments for a given field become an ordered set of zeroes and ones (or equivalently $+$ or $-$ $\mathbb{Z}_2$ charges) with $p$ elements, that we refer to as a charge vector below. This is especially convenient for computational representation and manipulation.  

We focus on the quark sector of the Standard Model which consists of three generations of left-handed doublets $Q_{L,i}$ and right handed singlets $u_{R,i}$ and $d_{R,i}$ for $i=1,2,3$.   Representing these fields generically by $X$, each has a charge vector $q(X)$ with binary components, one for each $\mathbb{Z}_2$ factor. It is convenient to regard these charge vectors as elements of the $p$-dimensional vector space $\mathbb{F}_2^p$, where $\mathbb{F}_2=\{0,1\}$ denotes the finite field with two elements,
\begin{equation}
q(X) \in \mathbb{F}_2^p \,\,\, .
\end{equation}
Addition of charge vectors is performed component-wise modulo 2, so that $1+1=0$. Consequently, invariance of an operator under the flavor symmetry is expressed simply by the vanishing of the sum of its constituent charge vectors in $\mathbb{F}_2^p$.  The Higgs field will be taken to be flavor neutral.

The discrete flavor symmetry will be spontaneously broken through vevs of $N_\phi$ flavon fields, $\phi_a$, which are singlets under the Standard Model gauge group. Each flavon field will also have binary charge assignments
\begin{equation}
q(\phi_a) \in \mathbb{F}_2^p 
\end{equation}
Trivial flavons (i.e., ones in which all charges are $0$) are excluded from the model, and all flavon charge vectors are required to be distinct.  Each flavon is assumed to have a vev, leading to the dimensionless ratios
\begin{equation}
\epsilon_a=\frac{\left<\phi_a \right>}{\Lambda_F}=\lambda^{r_a} \,\,\, ,
\label{eq:dimrats}
\end{equation}
where $\lambda=0.22$ is approximately the Cabibbo angle, a quantity that is known to be convenient for parameterizing Yukawa hierarchies~\cite{Wolfenstein:1983yz}, and $r_a$ is a positive integer.  For definiteness, we assume the range $1 \leq r_a\leq 8$, since higher powers are not needed to reproduce all the target observables, which typically depend on products/powers of the $\epsilon_a$.  Distinct flavons with equal exponents $r_a$ are associated with a common symmetry breaking scale and their collective action contributes to one step in the symmetry breaking chain shown in Eq.~(\ref{eq:chain}).  At each intermediate symmetry breaking scale, the unbroken flavor symmetry is $\mathbb{Z}_2^{p-R}$ where $R$ is the rank of the set of flavon charge vectors that have vevs at or above that scale.   If another flavon has a vev at a lower energy scale but its inclusion does not decrease $p-R$, then one cannot associate this vev with the breaking of a subgroup.  An anomalously small flavon vev of this type is not logically excluded but is disfavored by naive dimensional analysis. Here we follow the usual effective field theory power-counting arguments~\cite{Manohar:1983md,Georgi:1992dw} and assume higher-dimension operators are suppressed by appropriate powers of the heavy mass scale and have otherwise unsuppressed order-one dimensionless coefficients.  Models with flavon vevs that do not follow a sequential symmetry breaking pattern are penalized in our scan over theory space, as will be explained in Sec.~\ref{subsec:LF}.

To represent a given model, all the fermion and flavon charges may be collected in the matrices
\begin{equation}
    Q_f= 
    \begin{pmatrix}
        q(Q_{L,1})^T \\
        q(Q_{L,2})^T \\
        q(Q_{L,3})^T \\
        q(u_{R,1})^T \\
        q(u_{R,2})^T \\
        q(u_{R,3})^T \\
        q(d_{R,1})^T \\
        q(d_{R,2})^T \\
        q(d_{R,3})^T
    \end{pmatrix}
 \,\,\,\, \mbox{ and } \,\,\,\,
     Q_\phi =
    \begin{pmatrix}
        q(\phi_1)^T \\
        \vdots \\
        q(\phi_{N_\phi})^T
    \end{pmatrix} \,\,\, ,
\end{equation}
where $q(X)^T$ and $q(\phi_a)^T$ represent row vectors, each row corresponds to a distinct field and each column corresponds to a $\mathbb{Z}_2$ factor. A model is specified by the values of $p$ and $N_\phi$, the charges contained in the matrices $Q_{f}$ and $Q_\phi$, and the exponents $r=\{r_1, ..., r_{N_\phi}\}$ which give the ratios of flavon vevs to the cut off scale as powers of $\lambda$.

Yukawa couplings involve the gauge invariant field combinations $\overline{Q}_{L i} \widetilde{H} u_{R j}$ and $\overline{Q}_{L i} H d_{R j}$, in the up- and down-quark sector, respectively.  Invariance under $G_F$ will require that most of these operators also involve products of the flavon fields; the only exception is the $i=j=3$ operator in the up sector, which accounts for the heaviness of the top quark.   An invariant operator has the form
\begin{equation} 
 c_{ij}^{(f)} \, \overline{Q}_{Li} {\cal H} f_{Rj} \prod_{a=1}^{N_\phi} \left(\frac{\phi_a}{\Lambda_F}\right)^{n_{a,ij}^{(f)}} \,\,\, ,
\end{equation}
where ${\cal H}$ is $\widetilde{H}$ for $f=u$, and $H$ for $f=d$, the $c_{ij}^{(f)}$ are order-one coefficients, and the $n_{a,ij}^{(f)}$ are the flavon powers. Invariance follows from the condition 
\begin{equation}
     q(Q_{L,i})+q(f_{R, j})+\sum_{a=1}^{N_\phi} n_{a,ij}^{(f)} \, q(\phi_a)=0 \ \ \ \ \ (\text{mod} \ 2) \,\,\, .
\end{equation}
For each Yukawa entry, we retain the operator with the fewest flavon insertions, since higher-order operators are subleading. It is sufficient to take $n_{a,ij}^{(f)} \in \{0, 1\}$ since a pair of identical flavon insertions carries zero net charge and is of higher order than the omission of those fields.

The flavon expectation values generate elements of the Yukawa matrices,
\begin{equation}
   \left(Y_f\right)_{ij} = c_{ij}^{(f)} \prod_{a=1}^{N_\phi} \epsilon_a^{n_{a,ij}^{(f)}} \,\,\, ,
\end{equation}
where there is no sum on $i$ and $j$.  We then predict the following target observables:  (1) the six quark Yukawa couplings 
\begin{equation}
y_i \in \{y_u, y_c, y_t, y_d, y_s, y_b\}
\end{equation}
 which correspond to the eigenvalues of these matrices.  Quark masses are determined by the products $y_i \, v/\sqrt{2}$, where $v \approx 246$~GeV is the electroweak scale.  (2) The Cabibbo-Kobayashi-Maskawa (CKM) mixing angles, which are encoded in the values of $V_{us}$, $V_{ub}$ and $V_{cb}$.  For this feasibility study, we ignore the CKM phase and focus on the following target values that are qualitatively consistent with values renormalized at a typical grand unified scale, namely 
 \begin{equation}
\begin{array}{lclcl} 
y_u=\lambda^8 & \hspace*{1em} & y_d=\lambda^6 &\hspace*{1em} & |V_{us}|=\lambda \\
y_c=\lambda^4 & & y_s=\lambda^4 && |V_{cb}|=\lambda^2 \\
y_t=1 & & y_b=\lambda^2 && |V_{ub}|=\lambda^3 \,\,\, .
 \end{array} 
 \label{eq:targets}
 \end{equation}.  
In a realistic study, one would need to specify the assumed flavor scale, and take into account the renormalization of the Yukawa matrices (whose forms differ from theory to theory) between the flavor and the weak scale, to a desired loop order and with threshold effects included. While that will alter the target values from those shown in Eq.~(\ref{eq:targets}), we do not believe that this difference will significantly impact the qualitative assessment we presently seek of the usefulness of the search algorithms described in the next section.  In fact, these phenomenological steps add to the computational cost of exact model evaluation, which will make the machine learning algorithm discussed in Sec.~\ref{sec:ML} even more advantageous.

\section{Search Algorithms} \label{sec:ML}

In this section, we discuss our two methods of searching the space of possible $(\mathbb{Z}_2)^p$ models.   These were implemented via Python code that was written via prompt engineering using ChatGPT (GPT-6)~\cite{OpenAI}, with output checked directly for accuracy using 
Mathematica 15~\cite{Wolfram:Mathematica15}.

\subsection{The loss function} \label{subsec:LF}

Since our objective is to identify flavor structures that reproduce the observed hierarchies, rather than to perform a statistical fit weighted by experimental uncertainties, we quantify the discrepancy between predicted and target values using a weighted least-squares loss in logarithmic scale~\cite{Cowan:1998},

\begin{equation}\label{eq:losslog}
    \chi^2_{\log}(M,c)=
    \sum_k \dfrac{1}{\sigma_k^2}
    \left[
    \log\left(
    \frac{O_k^{\text{pred}}(M,c)}
         {O_k^{\text{target}}}
    \right)
    \right]^2,
\end{equation}
where $M$ denotes a specific model, $c$ represents the set of order-one coefficients, and the dimensionless parameters $\sigma_k$ determine the relative weights of the observables. This form measures multiplicative deviations from the targets on an equal footing across widely differing scales, so that over- and under-predictions by the same factor contribute equally to the loss function. When all predicted values are close to their targets, Eq.~(\ref{eq:losslog}) reduces, to leading order, to the conventional least-squares $\chi^2$, with $\sigma_k O_k^{\text{target}}$ playing the role of the one-standard-deviation uncertainty in the $k^{\text{th}}$ observable. For definiteness, we set $\sigma_k=0.1$ for all $k$; we explain later why this choice is not crucial for identifying promising models in our initial coarse-grained search.

Using Eq.~(\ref{eq:losslog}), both the over-prediction and under-prediction of the target values by a same factor will result in the same loss value. For each model, the best $\chi^2_{\text{best}}$ is chosen among the $\chi^2_{\text{log}}$ values for $N_\text{c}$ coefficient samples
\begin{equation}\label{eq:lossbest}
    \chi^2_{\text{best}} (M) = \min_{s=1,...,N_c} \ \chi^2_{\log} (M,c_s)
\end{equation}
where $\chi^2_{\text{best}}$ stands for the lowest value found within the finite scan.  When we use the term ``exact model evaluation" we refer to this discrete minimization over $N_\text{c}$ sets of model coefficients, which includes the diagonalization of the Yukawa matrices needed for the evaluation of Eq.~(\ref{eq:losslog}). Our full loss function $\chi^2$ includes complexity penalties that allow us to bias our results towards simpler models with desired symmetry breaking patterns:
\begin{equation}\label{eq:loss}
    \chi^2=\chi^2_{\text{best}} + \lambda_p \, p + \lambda_\phi \, N_\phi + \lambda_{\text{seq}} \, P_{\text{seq}} \,\,\, .
\end{equation}
Here the $\lambda$'s are adjustable penalty weights associated with each corresponding penalty term.   The second and third terms grow with the number of $\mathbb{Z}_2$ factors, $p$, and flavon fields, $N_\phi$, respectively.   The quantity $P_{\text{seq}}$ counts the number of times a distinct flavon vev does not correspond to a reduction in the quantity $p-R$, where $R$ is the rank of the set of flavon charge vectors that have vevs above that scale.  In other words, this penalty corresponds to a violation of the sequential symmetry breaking paradigm.  $P_{\text{seq}}$ also counts any $\mathbb{Z}_2$ factors that remain unbroken after all flavon vevs are taken into account, a less-than-minimal situation that we similarly disfavor.

\subsection{Active learning: random forest regressor} \label{subsec:AL}

Our active learning search uses a random forest regressor which predicts the value of the loss function of candidate models from the number of flavor symmetry factors and flavons, the powers of $\lambda$ appearing in the Yukawa textures, and diagnostics characterizing operator complexity and sequential flavor symmetry breaking. Starting from an initial dataset with exactly evaluated models, the random forest regressor is trained to predict the logarithm of the loss function Eq.~(\ref{eq:loss}). In each round, the regressor acts as a surrogate in place of exact model evaluation, predicting good models from a randomly generated candidate pool and exactly evaluating a designated number of the most promising models. The newly evaluated models are added to the training dataset and the process is repeated. In the end, the best models are selected to be refined further by the continuous optimization of their operator coefficients. The main steps of the active search algorithm are described in more detail below:

\begin{enumerate}

\item For definiteness, we choose the parameter ranges $p\in[2,6]$ and $N_\phi\in[1,\min(6,2^p-1)]$, where $2^p-1$ is the maximum number of distinct nonzero $(\mathbb{Z}_2)^p$ flavon charge assignments.\footnote{We omit $p=1$, which is unlikely to yield viable models.} An initial set of $N_\text{m}$ models is generated by randomly selecting $p$ and $N_\phi$ within these ranges and randomly assigning the field charges and flavon vevs. The Yukawa matrices are then constructed by identifying the lowest-order operators contributing to each matrix element. Each model is evaluated by scanning over randomly generated complex operator coefficients with $|c_{ij}^{(f)}|\in[0.9,1.1]$ and $0\leq\mathrm{Arg}\,c_{ij}^{(f)}<2\pi$. The model features and the mean, median, and best values of the total loss $\chi^2$ are recorded. 
    
The narrow range of operator-coefficient magnitudes is intentional, and is designed to avoid identifying models in which a conspiracy of coefficient choices compensates for powers of $\lambda$. In other words, we seek models in which as much of the flavor structure as possible is attributable to the $G_F$-breaking pattern rather than the choice of nuisance parameters. After identifying the most promising models through this coarse-grained approach, we subsequently minimize $\chi^2$ by continuously varying the operator coefficients over a wider range of magnitudes (see item 6 below).    

\item A random-forest regressor~\cite{Breiman:2001hzm}, implemented using scikit-learn~\cite{scikitlearn}, is trained on the exactly evaluated models to predict $\log_{10}(\chi^2_{\mathrm{best}})$ from their features. The forest consists of 300 decision trees, each trained on a bootstrap sample drawn with replacement from the available dataset.\footnote{Bootstrap sampling involves constructing a new training dataset by randomly drawing models from the original dataset with replacement, so that some models may appear multiple times while others are omitted. Each decision tree is trained on a separate such sample.} Each tree recursively partitions
the feature space and assigns to a candidate model the mean target value of the training examples in
the corresponding terminal leaf. The forest prediction is the average of the individual tree predictions. All available exactly evaluated 
models are used for training, without reserving a separate test set. 

\item  A large pool of $N_\text{c}$ candidate models is randomly generated. Their features are computed, but the models
are not exactly evaluated; in particular, no operator coefficient scans are performed at this stage.

\item The trained random forest predicts $\log_{10}(\chi^2_{\mathrm{best}})$ for each candidate using only its features. The $N_{\mathrm{active}}$ candidates with the lowest predicted losses are selected for exact evaluation, using the coefficient-scanning procedure described in item~1. The resulting evaluated models are added to the training dataset, and the forest
is retrained on the enlarged dataset. This process allows information from previous evaluations to guide
subsequent exploration of theory space.

\item
Steps 2--4 are repeated for $N_{\mathrm{rounds}}$ active-learning rounds, yielding a total of $N_\text{m}+N_{\mathrm{rounds}}N_{\mathrm{active}}$ exactly evaluated models, including the initial random sample.

\item
Following the coarse-grained search, we select the $N_{\mathrm{top}}$ models with the lowest total losses for further refinement. The discrete model parameters, including the field and flavon charge assignments and flavon VEVs, are held fixed while the complex operator coefficients are optimized using the L-BFGS-B algorithm~\cite{l-bfgs-b}. This gradient-based method approximates the local curvature of the loss function while respecting bounds on the optimization parameters. At this stage, the coefficient magnitudes are allowed to vary over the broader interval $|c_{ij}^{(f)}|\in[1/3,3]$. Multiple optimization restarts are employed to reduce sensitivity to local minima.

We also assess the local sensitivity of each refined model by randomly perturbing the optimized coefficient magnitudes and evaluating the resulting changes in $\chi^2$ without further optimization. In the implementation considered here, 20 independent perturbations are applied, with multiplicative log-normal fluctuations of width $0.10$; the coefficient phases are held fixed.

\item
For comparison, a separate random search is performed with the same total budget of exact evaluations, $N_{\mathrm{random}}=N_\text{m}+N_{\mathrm{rounds}}N_{\mathrm{active}}$. The models are generated and evaluated using the same parameter ranges and coarse-grained scoring procedure, but without use of the regressor as a surrogate.
        
\end{enumerate}

\subsection{Genetic search} \label{subsec:GA}

The genetic algorithm evolves a population of flavor models by repeatedly selecting, combining, and mutating the most successful candidate models.   The algorithm alternates between two distinct stages: (1) {\it evaluation}, in which the value of the loss function is computed for each candidate model by performing an exact coefficient scan, and (2) {\it evolution}, in which the current population is transformed into a new generation through {\it elite preservation}, {\it tournament selection}, {\it crossover} and {\it mutation}.  These four terms are explained below.  Unlike the active-learning algorithm described in Sec~\ref{subsec:AL}, every candidate model generated during the genetic search is evaluated exactly; no surrogate  process is used.  The algorithm proceeds as follows:

\begin{enumerate}

\item An initial population consisting of $N_\text{m}$ randomly generated flavor models is constructed for fixed values of the hyperparameters $p$ and $N_\phi$.  Fixing these quantities ensures that every model has the same field content, charge-vector dimensionality, and number of flavon fields, allowing crossover operations to be defined unambiguously (see item 5 below).   Each model is generated by randomly assigning quark-field charges, distinct nontrivial flavon charges, and flavon vev hierarchies.  The hyperparameters $p$ and $N_\phi$ are held fixed throughout a single genetic search, after which the entire procedure is repeated for many different $(p,N_\phi)$ pairs.

\item  Yukawa matrices are constructed by determining the lowest-dimension effective operators permitted by the flavor symmetry and evaluating the corresponding suppression factors using the flavon vev hierarchies. For each model, the order-one coefficients are sampled randomly over the restricted magnitude range $0.9\le |c_{ij}|\le1.1$. This coarse-grained search emphasizes the suppression factors determined by the powers of $\lambda\approx 0.22$ appearing in the Yukawa textures. Continuous optimization over a broader coefficient range, $1/3\le |c_{ij}|\le3$, is performed only after the genetic search has identified the most promising discrete models.

\item Models are then ranked according to the best value of the loss function.\footnote{The median value obtained in the scan over the coefficients is used as a tie breaker.} A specified number of the best models from this ranked list, $N_{\mathrm{elite}}$, are copied unchanged into the next generation.   For example, if $N_{\mathrm{elite}}=4$, then the four best models out of $N_{\text m}$ are included in the offspring population.   The remaining members of the next generation are generated according to the following steps.

\item 
Tournament selection is used to choose the parents for each offspring. A subset of the current population of fixed size is selected at random, and the model with the lowest loss function within that subset is chosen as a parent.  For example, if the tournament size is set to $3$, then a random set of three models is selected out of the $N_\text{m}$ available, and the best model of the three is designated to be a parent.  The procedure is repeated independently to obtain a second parent. This process is carried out anew to provide two parents for every offspring until the next generation has been filled.   Each generation contains the same number of models as the initial population.

\item Given the selection of parent models, offspring are generated through crossover and mutation.  A crossover probability determines whether a new offspring is produced by independently inheriting the quark charges, flavon charges, and flavon hierarchy exponents from either parent. For example, the $(\mathbb{Z}_2)^p$ charges of the left-handed up quark in the offspring may be selected from one parent model, while those of the right-handed charm quark may be selected from the other. If crossover is not performed, one of the two parents is cloned at random.

\item  The offspring models are then subject to random mutation of quark field charges, flavon charges and the flavon vev hierarchy exponents, with a specified probability.
Mutation may occasionally produce an invalid flavor model—for example, one containing a flavon that is neutral under the flavor symmetry or that duplicates the charge assignments of another flavon.  Such models are repaired by randomly regenerating the offending flavon charges until a valid model is obtained.

\item  The elite models and offspring are combined to form the next generation of $N_\text{m}$ models. This population is sorted according to the loss function, and the evolutionary cycle is repeated until the desired number of generations has been completed.

\end{enumerate}

For every exact model evaluation, the complete set of model features and diagnostic quantities (such as the best and median loss functions, the predicted flavor observables, complexity measures, etc.) is recorded.  This makes it possible to plot the best loss achieved as a function of the cumulative number of exact model evaluations and to reconstruct the Yukawa textures of the highest-performing models. The most promising models from the coarse-grained search are then refined by continuous optimization of the order-one coefficients over the broader coefficient range, the same final step used in the active learning search of Sec.~\ref{subsec:AL}.

\section{Results} \label{sec:results}

In this section, we present results that shed light on the usefulness of the machine learning and genetic algorithms that were described in Sec.~\ref{sec:ML}.   The first issue we address is whether these computational approaches are better than simply evaluating a large number of randomly generated models.     Here, we define ``large" as a number whose computation time would run at most over ${\cal O}(10)$ hours, so that a human theorist could study a range of hyperparameter choices (for example, the weights of the various penalty terms in the loss function, different values of random number generator seeds, etc.), in a reasonable amount of time.   Those choices by themselves represent an enormous parameter space. The longest runs that we discuss below, which involved ${\cal O}(10^5)$ exact model evaluations, are large in this sense; however, this is small compared to the entirety of the theory space that we have defined. For definiteness, our algorithms scan over the ranges $2 \leq p \leq 6$ and $1 \leq N_\phi \leq N_\phi^{\rm max}$, where $N_\phi^{\rm max} \equiv  \min \left(6, 2^p-1\right)$. For these choices, the number of distinct models is 
\begin{equation}
\sum_{p=2}^6 \sum_{N_\phi=1}^{N_\phi^{\rm max}} \, 2^{(9+N_\phi)\, p} \, 8^{N_\phi} \sim {\cal O}(10^{32}) \,\,\, ,
\label{eq:nummodels}
\end{equation}
where the power of $2$ reflects the $p$ charge assignments on each of the quark and flavon fields, and the power of $8$ refers to our chosen discrete sampling of the ratios of flavon vevs to the cutoff scale [see the discussion after Eq.~(\ref{eq:dimrats})].  The right-hand-side of Eq.~(\ref{eq:nummodels}) is far larger than ${\cal O}(10^5)$; nonetheless, we will see that the active learning approach quickly finds successful models.  
\begin{figure}[t]
\centering
\includegraphics[width=0.45\textwidth]{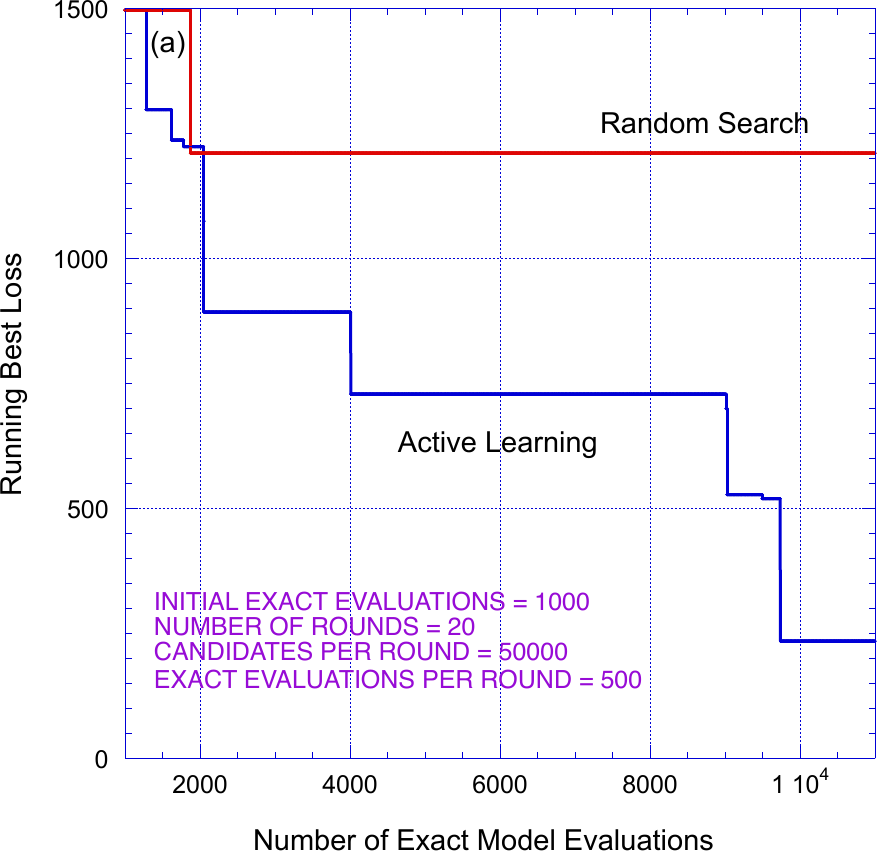} \hspace{1em}
\includegraphics[width=0.45\textwidth]{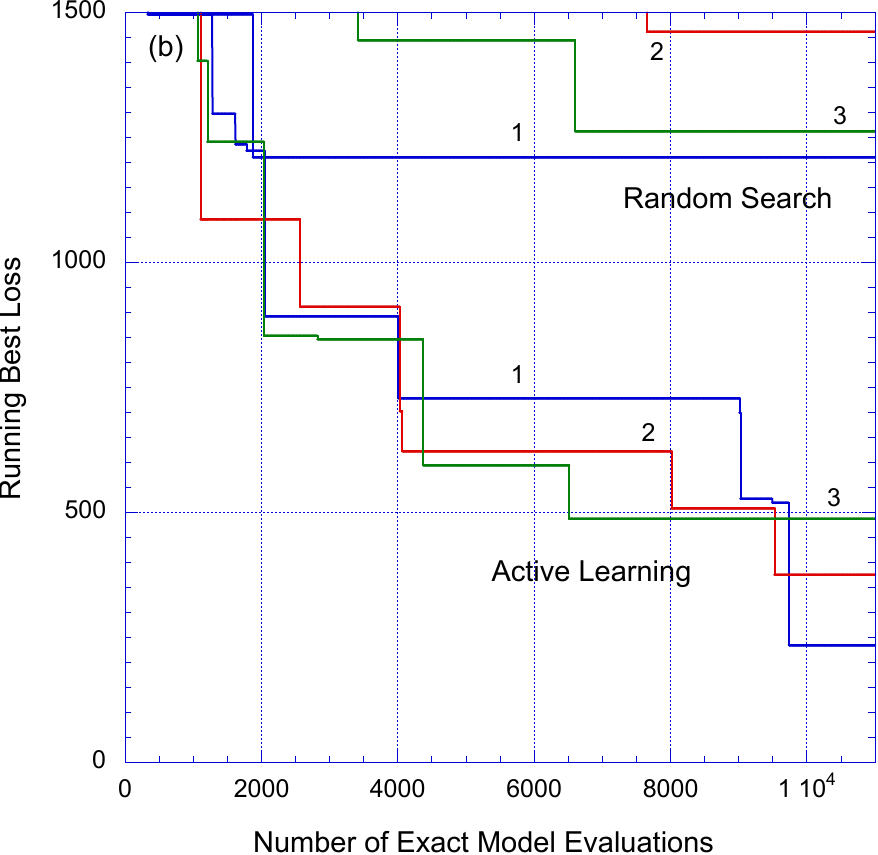}
\caption{Comparison of the best value of the loss function versus the number of exact model evaluations between the active learning approach and random model generation.  The inset in Fig.~\ref{fig:1}{\color{red}(a)} refers to hyperparameter choices described in the text.  The curves labeled $1$, $2$ and $3$ in Fig.~\ref{fig:1}{\color{red}(b)} show the results of varying the seeds for the random number generators used in the analysis.}
\label{fig:1}
\end{figure}

Fig.~\ref{fig:1} compares the best value of the loss function in a coarse grained active-learning search to what one obtains by studying models that are randomly generated.  In the example shown, the random forest regressor is trained on $1000$ randomly generated models that are exactly evaluated.  The regressor is applied as a surrogate to identify the $500$ best models out of $50,000$ models that are generated randomly in the next round.  Those best models are then exactly evaluated and added to the training set, so that the regressor actively learns how to identify better models.  In Fig.~\ref{fig:1}, the process is repeated a total of $20$ times after the evaluation of the initial training set.

With our choice of the $\sigma_i=0.1$, values of the loss function that are ${\cal O}(100)$ are consistent with textures that capture the desired mass eigenvalue and mixing angle hierarchies, in powers of $\lambda$.   With a coarse grained scan of coefficient choices, typically involving only $100$ randomly generated sets, we might expect predictions in the best models to match observables up to powers of $\lambda$, but not necessarily up to order one factors.  Deviation by a factor of $2$ from the target value of one observable, for example, contributes $\ln(2)^2/\sigma^2 \sim 48$ to the loss function; with $9$ observables that could receive similar contributions, an ${\cal O}(100)$ result is not surprising, even before taking into account the penalty terms in Eq.~(\ref{eq:loss}).   As a point of reference, a relatively simple model with no sequential penalty $P_{\text{seq}}=0$ and $p=4$ and $N_\phi=3$ would increase the total $\chi^2$ by $70$, for the choice $\lambda_p=\lambda_\phi=10$ assumed in Fig.~\ref{fig:1}, also consistent with the ${\cal O}(100)$ expectation.   This suggests that this choice of penalty weights will meaningfully bias the results towards simpler models.

The best models from the coarse grained search are improved as a final step through continuous minimization of the operator coefficients. As indicated earlier, the coarse-grained search restricts the range of each order-one parameter magnitude to be between $0.9$ and $1.1$.  This helps the algorithm identify models in which most of the relevant flavor physics originates from the powers of $\lambda$ appearing in the Yukawa matrices, which reflects the underlying flavor symmetry and its breaking pattern, rather than larger adjustments of the operator coefficients.  This aligns with the motivations of the sequential flavor symmetry breaking paradigm.    Continuous optimization of the operator coefficients typically leads to values of the penalty-free part of the $\chi^2$, i.e., Eq.~(\ref{eq:lossbest}), that are much smaller than the number of parametric degrees of freedom and sometimes close to zero, suggesting that smaller values of the $\sigma_i$ would affect the values of the loss in the coarse grained search, but not the optimized values.  However, this behavior hides that the coefficient values at the minimum of  Eq.~(\ref{eq:lossbest}) are often unnatural, either because of tunings between coefficients or because a significant fraction of the full set live near the endpoints of the interval $[1/3,3]$ that we take as representative of natural values.  We discuss how to identify models that are free of these problems later in this section.

As noted earlier, the penalty weights $\lambda_p$ and $\lambda_\phi$ are both set to $10$ in the example  shown in Fig.~\ref{fig:1}, to bias the results towards simpler models.  The remaining weight,  $\lambda_{\text{seq}}$, is set to $1000$.  The much larger value of the latter penalty is intended to push any model that violates the sequential symmetry breaking assumptions (i.e., that each smaller flavon vev is associated with a distinct subgroup of $G_F$) outside of the set of ``good" models, defined by choosing an appropriate upper limit on the value of the loss function, which we call $\chi^2_{\text{max}}$.   There are clearly an infinite number of possible plots one could make depending on the choice of the penalty weights; in some sense this parallels the continuum of subjective aesthetic choices that a model builder makes in deciding whether a model is ``simple enough".   Nevertheless, we believe the example shown gives a qualitative sense of the results that we would obtain using other reasonable choices.   

\begin{figure}[t]
\centering
\includegraphics[width=0.5\textwidth]{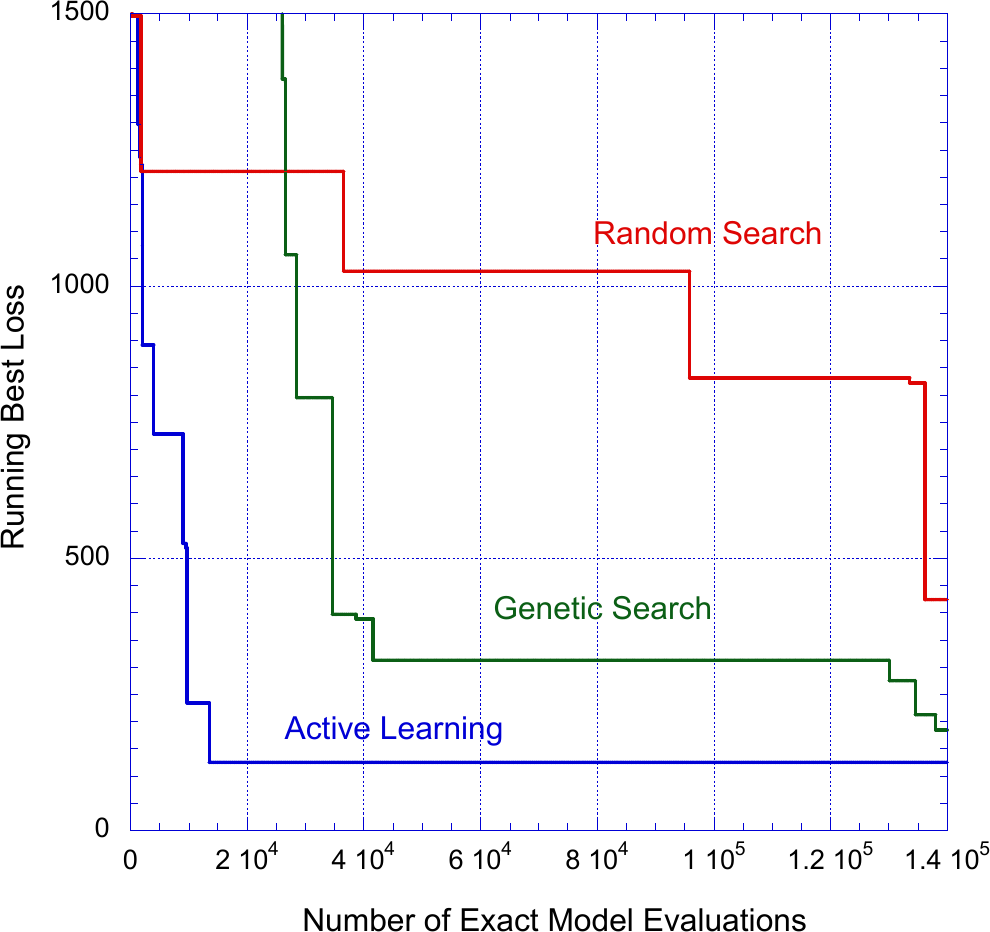}
\caption{Comparison of the best value of the loss function versus the number of exact model evaluations, like Fig.~\ref{fig:1}, but now also including the results of the genetic algorithm.   The random seed choices are the same as in the left subfigure of  Fig.~\ref{fig:1}.}
\label{fig:2}
\end{figure}

The plot in Fig.~\ref{fig:1}{\color{red}b} involves the same hyperparameter choices as Fig.~\ref{fig:1}{\color{red}a}, but with different selections of the random seeds used to define the models and the regressor.   Within our code, a variable named {\it MODEL\_SEED} controls the random generation of candidate flavor models, including the choices of $p$, $N_\phi$, quark and flavon charge assignments, and flavon vev hierarchies. The variable {\it COEFFICIENT\_SEED} controls the random sampling of the complex order-one Yukawa coefficients (magnitudes and phases) used to evaluate the $\chi^2$ of each candidate model.  Finally,  {\it ML\_SEED} controls the random processes involved in constructing the random-forest regressor, principally the bootstrap sampling of training models used to build its individual decision trees.  In Fig.~\ref{fig:1}, the seeds are taken to be strings of 5 non-negative integers, and these are all different in the curves labeled by $1$, $2$ and $3$ in the plot shown on the right. Fig.~\ref{fig:1} illustrates both the effectiveness of the active learning search relative to random model generation, and that differences stemming from the random processes that are central to the model generation and the regressor do not change the qualitative conclusions.

One can compare these results with the evolution of the best value of the loss function versus number of exact model evaluations in the genetic algorithm described in Sec.~\ref{subsec:GA}.  This is shown in Fig.~\ref{fig:2}.   The genetic algorithm clearly outperforms random model generation, but requires many more model evaluations to achieve results similar to the active learning approach. Notice that the abscissa now extends to 140,000; plotted over the more limited range shown in Fig.~\ref{fig:1}, the genetic algorithm curve would not be visible in the plot.  

\begin{figure}[t]
\centering
\includegraphics[width=0.42\textwidth]{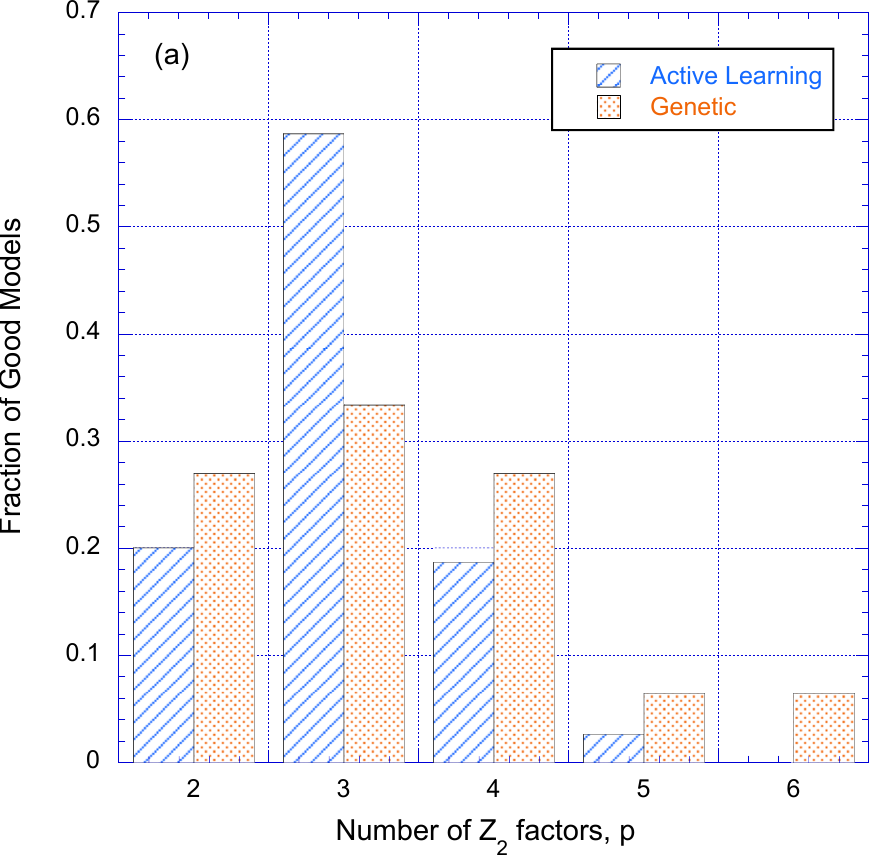} \hspace{1em}
\includegraphics[width=0.42\textwidth]{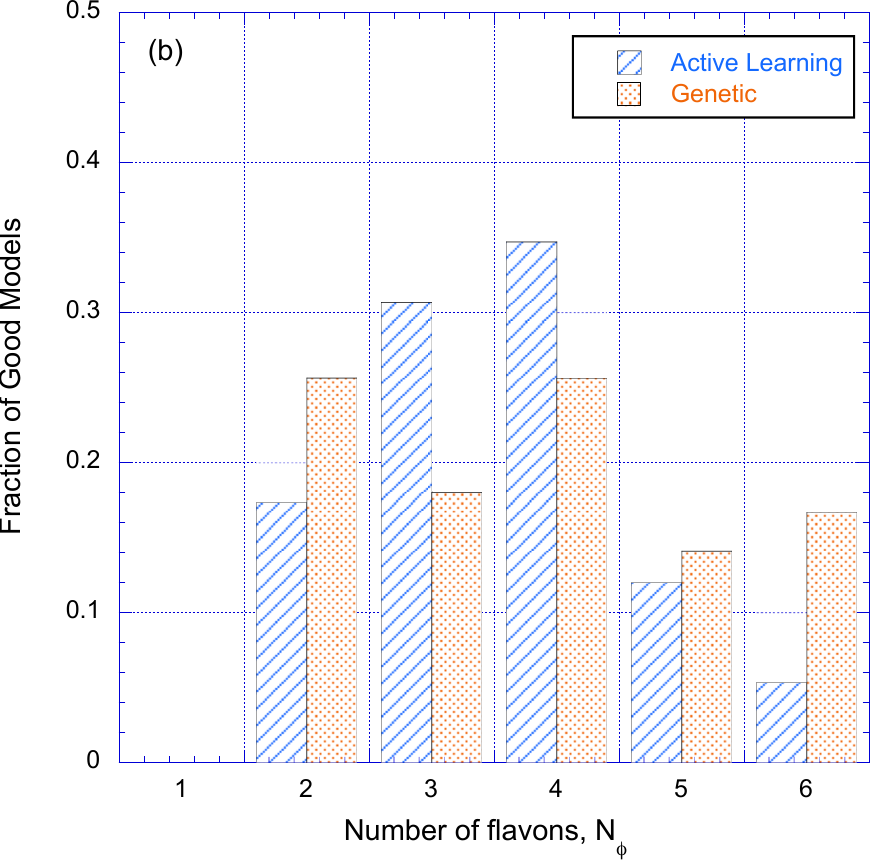} 
\caption{Distribution of models with loss function less than $\chi^2_{\text max}=500$ versus $p$ in Fig.~\ref{fig:3}{\color{red}(a)} and $N_\phi$ in Fig.~\ref{fig:3}{\color{red}(b)}.}
\label{fig:3}
\end{figure}
Our computational approach also allows us to make statistical observations of the properties of good models.  Setting $\chi^2_{\text{max}}=500$ henceforth, we show in Fig.~\ref{fig:3} the distribution of models with $\chi^2 < \chi^2_{\text{max}}$, as a function of $p$ and $N_\phi$, respectively,  assuming only a large penalty coefficient (equal to $1000$) for the sequential penalty term in the loss function.   This restricts the population of models to those which are consistent with the sequential symmetry breaking paradigm, without biasing the search in these particular plots against large values of $p$ and $N_\phi$.   As one might expect, the distributions show fewer good models at small values of $p$ and $N_\phi$ where models are less likely to successfully predict the complete set of target values.    The smaller histogram bars at larger $p$ and $N_\phi$ are likely due to a combinatorial suppression:  there are a larger number of possible models for larger fixed values of these hyperparameters, so the probability that a good model will be randomly generated for these choices when the integrated total is held fixed should be reduced.    We don't dwell on this point since we are ultimately interested in biasing the loss function to disfavor large values of $p$ and $N_\phi$.   The plots illustrate that there are a range of reasonably small values of $p$ and $N_\phi$ (roughly between $2$ and $4$) where the coarse grained search finds models whose loss function value is consistent with what one would expect from a viable hierarchy in powers of $\lambda$.

\begin{figure}[t]
\centering
\includegraphics[width=0.42\textwidth]{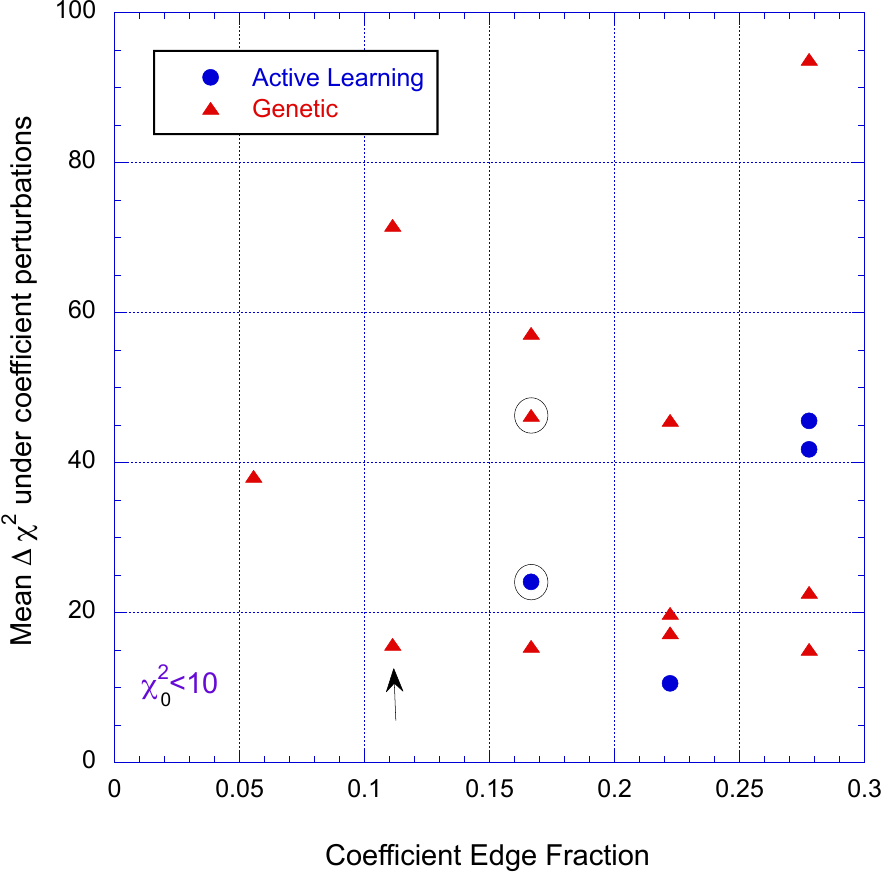} 
\caption{Model desirability after coefficient optimization, where $\chi^2_0$ is the optimized value of the loss without penalties.  Models with the least coefficient tuning, and the smallest number of coefficients near the edge of the range suggested by naive dimensional analysis fall towards the lower-left corner of the plot.  The circled points represent the two models discussed in Sec.~\ref{sec:xMLmodel} and \ref{sec:xgmodel}.  Note that the point indicated by an arrow has $p=4$ and $N_\phi=5$, and is less minimal than the models selected as examples in these sections.}
\label{fig:4}
\end{figure}

Returning to the models generated with the penalty coefficients and parameter choices that led to Fig.~\ref{fig:2}, and restricting our attention to models with a loss function less than $\chi^2_{\text{max}}$, we perform a continuous minimization of the ${\cal O}(1)$ operator coefficients, restricting them to be in a range that is consistent with naive dimensional analysis, with magnitudes between $1/3$ and $3$.    It is not difficult to find models in which the target observables can be exactly matched by coefficient choices that are returned by this optimization procedure.  However, this does not imply that these models are natural.   For example, continuous minimization can lead to fine tunings between the operator coefficients;  on the other hand, even if fine tuning is absent, a large number of coefficients may lie at the extreme edges of the preferred range between $1/3$ and $3$, and may conspire so that observables are much larger or much smaller than what would be expected from $\lambda$ power counting.   From an aesthetic perspective, a model builder would not likely be content with a Yukawa matrix that had, for example, six out of nine operator coefficients with magnitudes of $2.9$.   A smaller number of models remains when we exclude those models that suffer from either or both of these problems, depending on how much of each we are willing to tolerate.  This is illustrated in the scatter plot shown in Fig.~\ref{fig:4}.
The $x$-axis of the plot shows the ``coefficient edge fraction," which is the fraction of the total number of operator coefficients whose magnitudes are close to the extreme values suggested by naive dimensional analysis.   For definiteness, we define ``close" by
\begin{equation}
|c_a|<0.4 \,\,\,\,\, \mbox{ or }\,\,\,\,\, |c_a|>2.5. \,\,\,.
\end{equation}
The $y$-axis shows the mean variation in the loss function as we perturb about the best-fit parameter values.   This quantity is determined in the following way:  For each model, we generate $20$ independent perturbations in which the magnitude of each nonzero coefficient is multiplied by a factor $\exp(\delta)$ where $\delta$ is drawn independently from a Gaussian distribution with zero mean and standard deviation $0.10$.  This is based on the requirement that a good model will not have large changes in the total $\chi^2$  when the coefficient values vary by ${\cal O}(10\%)$. The perturbed magnitudes are restricted to the original optimization interval $[1/3,3]$, while the complex phases are held fixed. We then reevaluate the total loss without reoptimizing the coefficients and calculate the change relative to its value at the best-fit point. The median of these $20$ changes provides a measure of the local sensitivity of the fit to variations in the coefficient magnitudes. Smaller values indicate greater stability under the specified perturbations, while larger values indicate a stronger dependence on the precise coefficient values.   More desirable models therefore will fall closer to the lower-left corner of the plot shown in Fig.~\ref{fig:4}.  

Both of our search methods identify successful models.   We now present an example for each, corresponding to the circled data points in Fig.~\ref{fig:4}.

\subsection{Example model - active learning search} \label{sec:xMLmodel}
The following model was found in the active learning search that led to Fig.~\ref{fig:1}: this involved an initial training set of $1000$ models, and the subsequent iterative generation of 50,000 random models with 500 chosen by the random forest surrogate for exact evaluation in each loop and inclusion in the training set.

The flavor symmetry and number of flavon fields is given by
\begin{equation}
G_F = (\mathbb{Z}_2)^3 \,\,\,\,\,\mbox{,}\,\,\,\,\, N_\phi=4 \,\,\, ,
\end{equation}
and the field charge assignments are
\begin{equation}
\begin{array}{lcrcr} 
\multicolumn{5}{c}{\mbox{Fermions:}} \\ \hline\hline
Q_{1L} \sim (-,-,+) & & u_R \sim (-,-,-) & & d_R \sim (+,-,+)\\
Q_{2L}  \sim (+,-,-) &  & c_R \sim (-,+,+) & & s_R \sim (-,+,+)\\
Q_{3L}  \sim (+,+,+)& & t_R \sim (+,+,+) & & b_R \sim (-,+,-) \\ 
\hline\hline
\end{array}
\end{equation}
\begin{equation}
\begin{array}{lcrcrcr} 
\multicolumn{7}{c}{\mbox{Flavon Charges:}} \\ \hline\hline
\phi_1 \sim (-,+,-) & & \phi_2 \sim (-,-,-) & & \phi_3 \sim (-,-,+) & & \phi_4 \sim (+,-,+) \\
\langle \phi_1 \rangle/\Lambda_F = \lambda^2 && \langle \phi_2 \rangle/\Lambda_F = \lambda^4 
&& \langle \phi_3 \rangle/\Lambda_F = \lambda^4 & &  \langle \phi_4 \rangle/\Lambda_F = \lambda^4 \\
\hline\hline
\end{array}
\end{equation}
where we also show the flavon vevs. The Standard Model Higgs field is a flavor singlet, $H \sim (+,+,+)$.  Notice that the two distinct flavon vev scales correspond to the
sequential symmetry breaking
\begin{equation}
(\mathbb{Z}_2)^3 \stackrel{\langle \phi_1 \rangle}{\xrightarrow{\hspace{0.5cm}}} (\mathbb{Z}_2)^2
\stackrel{\langle \phi_2 \rangle\ldots\langle \phi_4 \rangle }{\xrightarrow{\hspace{1.2cm}}} \mbox{nothing} \,\,\, ,
\label{eq:MLchain}
\end{equation}
so that no sequential penalty is incurred when the model is evaluated.  The resulting quark Yukawa textures are
\begin{equation}
Y_u \sim \left(\begin{array}{ccc} \lambda^{8} & \lambda^{4} & \lambda^{4} \\ \lambda^{8} & \lambda^{4} & \lambda^{6} \\ \lambda^{4} & \lambda^{8} & 1\end{array}\right)  \,\,\, \mbox{ and } \,\,\,
Y_d \sim \left(\begin{array}{ccc}\lambda^{8} & \lambda^{4} & \lambda^{6} \\ \lambda^{8} & \lambda^{4} & \lambda^{4} \\ \lambda^{4} & \lambda^{8} & \lambda^{2}\end{array}\right) \, ,
\label{eq:mlxtex}
\end{equation}
with complex operator coefficients
\begin{equation}
C_u=\left(\begin{array}{ccc}-0.3012 - 0.6378\,\mathrm{i} & -0.2982 - 0.149\,\mathrm{i} & 0.9993 - 0.2497\,\mathrm{i} \\ -0.6942 + 0.9945\,\mathrm{i} & 0.7665 - 0.549\,\mathrm{i} & -2.813 + 0.8659\,\mathrm{i} \\ -0.2126 - 0.4717\,\mathrm{i} & -0.6214 - 2.78\,\mathrm{i} & 0.04529 + 0.999\,\mathrm{i}\end{array}\right) \,\,\, ,
\end{equation}
\begin{equation}
C_d=\left(\begin{array}{ccc}0.419 - 0.1344\,\mathrm{i} & -0.01502 + 0.4839\,\mathrm{i} & 0.2248 - 1.787\,\mathrm{i} \\ -0.4312 + 0.2341\,\mathrm{i} & -0.8642 - 0.09336\,\mathrm{i} & 0.901 + 0.4792\,\mathrm{i} \\ -0.01231 + 2.406\,\mathrm{i} & -0.5273 + 0.5485\,\mathrm{i} & -0.7012 - 0.7016\,\mathrm{i}\end{array}\right)\,\,\, .
\end{equation}
(Note that our code does not implement phase redefinitions to place these coefficient matrices into any particular desired form.) The coarse-grained search that identified this model, with textures in Eq.~(\ref{eq:mlxtex}), had the total loss $\chi^2=234.2$. Continuous complex coefficient optimization can lead to an exact matching of the nine target observable values, with a coefficient edge fraction of $0.167$, and a mean $\Delta \chi^2 = 24.1$ found by the coefficient perturbation described earlier in this section.  

\subsection{Example model - genetic search} \label{sec:xgmodel}

The following is a successful model produced by the genetic search.\footnote{For each choice of $p$ and $N_\phi$, the genetic search used a population of 500 candidate models evolved over 50 generations, with five elite models retained per generation and tournament selection among three candidates. Row-wise uniform crossover was applied with probability 0.9, while mutation probabilities were 0.1 for quark-field charge vectors, 0.25 for flavon charge vectors, and 0.25 for flavon VEV exponents.}  Like the example in Sec.~\ref{sec:xMLmodel}, $p=3$, but this one involves
a smaller number of flavon fields:
\begin{equation}
G_F = (\mathbb{Z}_2)^3 \,\,\,\,\,\mbox{,}\,\,\,\,\, N_\phi=3
\end{equation}
\begin{equation}
\begin{array}{lcrcr} 
\multicolumn{5}{c}{\mbox{Fermion Charges:}} \\ \hline\hline
Q_{1L} \sim (-,+,-) & & u_R \sim (+,-,-) & & d_R \sim (+,-,-)\\
Q_{2L}  \sim (-,+,-) &  & c_R \sim (-,-,-) & & s_R \sim (-,-,-)\\
Q_{3L}  \sim (+,+,+)& & t_R \sim (+,+,+) & & b_R \sim (+,+,-) \\ 
\hline\hline
\end{array}
\end{equation}
\begin{equation}
\begin{array}{lcrcr} 
\multicolumn{5}{c}{\mbox{Flavon Charges and vevs:}} \\ \hline\hline
\phi_1 \sim (+,+,-) & & \phi_2 \sim (+,-,-) & & \phi_3 \sim (-,+,+) \\
\langle \phi_1 \rangle/\Lambda_F = \lambda^2 && \langle \phi_2 \rangle/\Lambda_F = \lambda^2 
&& \langle \phi_3 \rangle/\Lambda_F = \lambda^4 \\
\hline\hline
\end{array}
\end{equation}
The Standard Model Higgs field is a flavor singlet, $H \sim (+,+,+)$.  Notice that the two distinct flavon vev scales correspond to the
sequential symmetry breaking
\begin{equation}
(\mathbb{Z}_2)^3 \stackrel{\langle \phi_1 \rangle,\langle \phi_2 \rangle}{\xrightarrow{\hspace{1.2cm}}} \mathbb{Z}_2
\stackrel{\langle \phi_3 \rangle}{\xrightarrow{\hspace{0.5cm}}} \mbox{nothing} \,\,\, ,
\label{eq:xchain}
\end{equation}
so that no sequential penalty is incurred when the model is evaluated.  The resulting quark Yukawa textures are
\begin{equation}
Y_u \sim \left(\begin{array}{ccc}\lambda^{8} & \lambda^{4} & \lambda^{6} \\ \lambda^{8} & \lambda^{4} & \lambda^{6} \\ \lambda^{2} & \lambda^{6} & 1\end{array}\right)
 \,\,\, \mbox{ and } \,\,\,
Y_d \sim \left(\begin{array}{ccc} \lambda^{8} & \lambda^{4} & \lambda^{4} \\ \lambda^{8} & \lambda^{4} & \lambda^{4} \\ \lambda^{2} & \lambda^{6} & \lambda^{2}\end{array}\right) \, ,
\label{eq:gxtex}
\end{equation}
with complex operator coefficients
\begin{equation}
C_u=\left(\begin{array}{ccc}1.117 - 0.09552\,\mathrm{i} & 0.6659 + 0.2177\,\mathrm{i} & -1.266 - 0.5099\,\mathrm{i} \\ -0.09493 + 0.7085\,\mathrm{i} & -0.5458 - 0.4596\,\mathrm{i} & 1.379 - 0.2191\,\mathrm{i} \\ -1.16 - 0.09368\,\mathrm{i} & 0.6121 - 0.6467\,\mathrm{i} & 0.9916 + 0.1164\,\mathrm{i}\end{array}\right) \,\,\, ,
\end{equation}
\begin{equation}
C_d=\left(\begin{array}{ccc}-0.5523 + 0.2743\,\mathrm{i} & -0.03983 + 0.578\,\mathrm{i} & -0.6409 - 0.8277\,\mathrm{i} \\ -1.102 - 0.547\,\mathrm{i} & 0.1681 - 0.3186\,\mathrm{i} & 0.2947 + 0.639\,\mathrm{i} \\ 0.3587 + 0.4562\,\mathrm{i} & -0.4865 + 0.3679\,\mathrm{i} & 0.6291 + 0.5148\,\mathrm{i}\end{array}\right)
\,\,\, .
\end{equation}
The coarse-grained search that identified this model, with textures in Eq.~(\ref{eq:gxtex}), had the total loss $\chi^2=184.0$. Continuous complex coefficient optimization can lead to an exact matching of the nine target observable values, with a coefficient edge fraction of $0.222$ and a mean $\Delta \chi^2 = 45.6$ found by our coefficient perturbation analysis.  

\section{Conclusions} \label{sec:conc}

In this work, we have considered two approaches to automating the search for models of flavor based on discrete flavor symmetries.  We have focused on the flavor group $G_F=\left(\mathbb{Z}_2 \right)^p$ in models with $N_\phi$ flavon fields; these models are easy to represent and randomly generate using Python code.   We considered how the space of possible models could be searched using two methods: an active learning approach that employs a random forest regressor, and a genetic search.   The exact evaluation of an equivalent number of randomly generated models was used as a control for comparison in both cases.   After appropriate training, the random forest regressor is used as an effective surrogate for exact evaluation in identifying the most promising models of a large, randomly generated set; this subset is subsequently evaluated exactly and added to the training set.  Focusing the exact evaluations on a subset of promising candidates allows more efficient use of computational effort. Through active learning, the algorithm becomes better able to identify promising models as the search progresses, but does not create new models based on the features of the best models it has previously identified.   The genetic search, on the other hand, creates better models through a process of evolutionary optimization, selecting features of new models based on those of preferred candidates from a previous generation.  Both the active learning and genetic  approaches identified models with considerably lower losses than the random search for the same number of exact model evaluations.  Between the two, the active learning search reached the lowest loss models more quickly, after fewer exact evaluations.  These qualitative results were reproduced using different random number generator seeds that are necessary for model generation and  the functioning of the random forest regressor, indicating the robustness of our results.  The large numbers of models generated in an automated search also allowed us to consider the statistical properties of the features of good models, and to bias the results towards simpler models by assigning complexity penalties to the loss function used in the model evaluation.

In our analysis, exact model evaluation involved a random scan over the numerical values of the operator coefficients that contribute to the up- and down-quark Yukawa textures, diagonalization of those matrices, identification of the flavor observables, and then the determination of the mean, median and best values of the $\chi^2$ function over the coefficient sampling.   In a fully realistic analysis, exact model evaluation would be even more computationally intensive since determination of the $\chi^2$ function would also entail the numerical solution of the renormalization group equations to match model predictions to target values that are defined at much lower, experimentally accessible energy scales.  This suggests that the approach of using the random forest regressor as a surrogate for exact evaluation may provide a greater advantage in that context.  The genetic approach presented here may be disadvantaged since every generated model is exactly evaluated.  However, it should be possible to combine these approaches and use a trained regressor within the genetic algorithm to reduce the computational effort.  We reserve this possibility for future work.  

Our discrete operator coefficient scan provides for a coarse-grained search which we have shown allows us to identify Yukawa textures that are successful at the level of $\lambda$ power counting, where $\lambda \approx 0.22$ is the Cabibbo angle.   We then take the best models from the coarse-grained search and perform a continuous optimization of the coefficients, while keeping the other model features fixed; this reduces the loss more quickly than a much larger discrete scan over the space of possible coefficient values.  However, as a consequence of the number of operator coefficients that are present in models with purely Abelian discrete flavor symmetries, the question is not whether a continuous optimizer can find values of the coefficients that reproduce all the desired target observables, but whether it can do so consistent with naive dimensional analysis and free from any unnatural correlations or  fine-tunings.   We studied the best models from our coarse-grained searches, numerically evaluating the mean change in the loss function after a series of controlled perturbations of the operator coefficients about their best values, to reveal sensitivity that would indicate fine-tuning.  We also evaluated the fraction of the total number of operator coefficients whose magnitudes are near the boundaries of the interval $[1/3,3]$ that we use to define agreement with naive dimensional analysis.   We find that a small fraction of the models have both low sensitivity to perturbations and small numbers of coefficients with magnitudes near the boundaries of the preferred range.   Examples of such models from our active learning and genetic searches were presented in Sec.~\ref{sec:results}.

Our present study is limited by our specification of the theory space of interest, our choices for hyperparameter ranges and target observables, and our focus on algorithmic effectiveness rather than phenomenological detail.  The ability of the active learning and genetic algorithms to find models that reproduce the hierarchy in target observables that we have assumed is a necessary preliminary result that motivates a comprehensive analysis of more realistic models.  In particular, the present analysis may be extended to the charged leptons and the neutrino sector, with the CP-violating phase in the CKM matrix included;  renormalization group running and accurate low-energy flavor data can be taken into account; and the constraints from the cancellation of discrete gauge anomalies may lead to further interesting restrictions.   These issues, and the possible inclusion of a non-Abelian factor, 
will be considered in follow-up work.

%%%%%%%%%%%%%%%%%%%%%%%%%%%%%%%%%%%%%%%%%%%%%%%%%%%%%%%%%%%
\newpage
\begin{acknowledgments} 
We thank the NSF for support under Grant No. PHY-2411549.
\end{acknowledgments}

%\appendix

\end{document}